# Optical outbursts of the cataclysmic variable 1RXS J140429.5+172352

Jeremy Shears, Pavol Dubovský, Colin Littlefield, Ian Miller, Etienne Morelle, Roger Pickard, Javier Ruiz and Richard Sabo

**Abstract**

We present the nine-year light curve of the cataclysmic variable 1RXS J140429.5+172352 from 2005 April to 2014 July. We identified four dwarf nova-like outbursts, which typically lasted 3 to 5 days, with an amplitude of 3.1 to 5.3 magnitudes. Time resolved photometry during two outbursts revealed small hump-like structures with peak-to-peak amplitude up to 0.5 mag. They occurred on timescales of 20 min to an hour, but did not exhibit a stable period. We suggest that the system might be an Intermediate Polar.

**Introduction**

1RXS J140429.5+172352 was identified as a bright X-ray source in the ROSAT catalogue (1) which was likely an Active Galactic Nucleus (AGN) (2) with an optical counterpart (3). The X-ray flux was 0.140(022) counts/s, with hardness ratio HR 1= 0.40(15), HR 2 = 0.38(18). The Sloan Digital Sky Survey (SDSS) spectrum of the 17$^{th}$ magnitude optical counterpart (SDSS J140429.37+172359.4 (4), with u=17.11, g=17.49, r=17.36, i=17.38, z=17.26) shows strong single-peaked line profiles of Balmer emission along with a weaker He I emission (5) and is indicative of a Cataclysmic Variable (CV) rather than an AGN. Moreover the redshift measured by SDSS is near zero (5), consistent with an object in the galaxy, which clearly rules out an AGN, which are rapidly receding extragalactic objects. The object has counterparts in the near-infrared, 2MASS J14042935+1723599 (6), and in the near-ultraviolet, GALEX J140429.3+172359 (7).

CVs are semi-detached binaries in which a white dwarf primary accretes material from a secondary star via Roche lobe overflow. The secondary is usually a late-type main-sequence star. In the absence of a significant white dwarf magnetic field, material from the secondary is processed through an accretion disc before settling on the surface of the white dwarf. In CVs with low to intermediate mass transfer rates, dwarf nova outbursts with amplitudes of 2–8 mag and durations of days to weeks are observed. These quasi-periodic outbursts are thought to be caused by a thermal instability in the accretion disc. Outburst recurrence times vary from system to system and depend primarily on the mass transfer rate from the donor star and the size of the accretion disc. In cases where the white dwarf has a stronger magnetic field, the inner portion of the disc becomes truncated and the accreting material follows the magnetic field lines and falls onto the poles of the white dwarf. Such systems are called Intermediate Polars (IPs). IPs also exhibit short-term variability mostly associated with turbulence in the accretion system. For more information about CVs, dwarf novae and IPs, the reader is directed to the books by Warner (8) and Hellier (9).

In this paper we report our analysis of nine years of photometry of 1RXS J140429.5+172352 and time-resolved photometry conducted during two outbursts, one in 2009 and one in 2014.

## Observations

*Long-term photometry*

To produce a long-term light curve of 1RXS J140429.5+172352, we used data from the AAVSO International Database, which contains more than 1800 observations of the star comprising both visual estimates and CCD measurements, and our own photometry. In addition we used 119 data points from the Catalina Real-Time Transient Survey (CRTS) (10), supplemented with a handful of outburst data from the MASTER-Amur survey (11) reported by Denisenko (12), ASAS-SN (the All Sky Automated Survey for Supernovae) (13) and ASAS-3 (All Sky Automated Survey) (14).

*Time resolved photometry*

We obtained 22.5 h of photometry during the 2009 outburst of 1RXS J140429.5+172352 and 28h during the 2014 outburst. Our instrumentation is shown in Table 1 and the observation log in Table 2. All photometry was unfiltered, except for RP's, which was obtained with a V-filter. Images were dark-subtracted and flat-fielded prior to being measured using differential aperture photometry relative to the BAA VSS preliminary V-band sequence P140520 based on APASS (AAVSO Photometric All Sky Survey) photometry (15).

## Results

*Long-term light curve*

The nine-year light curve of 1RXS J140429.5+172352, from 2005 April to 2014 July is shown in Figure 1. We identified four outbursts, taking an outburst as a brightening to magnitude 16.0 or brighter, the details of which are given in Table 3. The 2009 and 2014 outbursts lasted 3 to 5 days and the maximum observed brightness was magnitude 13.1 and 14.1 respectively. The other two outbursts were very poorly constrained, so it is not possible to conclude anything useful about their duration. The maximum recorded brightness in the 2008 outburst was 11.9 and in 2012 it was 14.2, although these might not represent the peak of the outburst, given the sparse data. The time between recorded outbursts was between 309 and 1000 days. However, given the relatively low cadence of the observations, the seasonal gaps in observations, and the relatively short duration of the observed outbursts, we note the light curve is under sampled and it is probable that we have missed some outbursts. The median interval between observations was 2 days and the mean was 5.8 days.

The mean quiescent magnitude was 17.2. There was considerable variation in quiescent brightness, with a standard deviation of 0.3 mag and a minimum of 18.1. Whilst some of this apparent scatter might be explained by errors inherent in the measurements (in the case of CRTS data, the mean error was 0.09 mag), the majority can be attributed to real variations in brightness.

Careful examination of Figure 1 also shows a possible long-term evolution in quiescence brightness. Fitting a third order polynomial to the quiescence data, i.e. having eliminated the outbursts and the "fainter than" observations (shown as the line in Figure 1 inset) to guide the eye shows this trend, although based on the available data it is only suggestive.

*The 2009 and 2014 outbursts*

Expanded light curves of the 2009 and 2014 outbursts are shown in Figure 2. Both outbursts were detected by JS as part of a long-term surveillance programme of this star. The 2009 outburst was detected on May 1.931 at magnitude 14.07 (unfiltered CCD with V zero point) (16) and was apparently on the final approach to outburst maximum which it reached later the same night. The outburst was short-lived, with a sharp maximum and had already returned to quiescence 3 days after it was detected. The amplitude above mean quiescence was 3.1 magnitudes.

The 2014 outburst was detected on May 17.954 at magnitude 14.8 (17) although it later transpired that ASAS-SN had detected the object a few hours earlier, on May 17.46 at V= 14.71 (13); the star had previously been observed in quiescence 7 days before that. The star continued to brighten gradually over the next 3 nights after which a rapid fade set in. The approach to quiescence was not well observed, but the overall outburst lasted at least 5 days. The amplitude above mean quiescence was 4.1 magnitudes.

*Time resolved photometry*

Figures 3 and 4 show some of the longer photometry runs obtained during the 2009 and 2014 outbursts plotted to the same scale. Throughout both outbursts small hump-like structures were observed. These were variable in size but the larger ones had a peak-to-peak amplitude of ~0.5 mag. Superimposed on these was small-scale flickering. We analysed the combined datasets for each outburst, as well as the daily time resolved photometry runs, using the Lomb-Scargle and ANOVA algorithms in the Peranso V2.50 software (18), but could not find a significant stable period in the power spectra. Visual inspection of the light curves shows that the humps occur in an irregular manner with an interval between them of 20 mins to about an hour.

*Astrometry*

We performed astrometry on our images of 1RXS J140429.5+172352 using the *Astrometrica* software (19) and the UCAC2 astrometric database (20) which yielded a mean position of 10 separate measurements of RA 14h 04min 29.36sec  Dec. 17° 23' 59.8" (J2000.0), with mean errors of 0.10 and 0.09 arcsec in RA and Dec respectively.

**Discussion**

The outburst behaviour we have observed in 1RXS J140429.5+172352 is consistent with normal outbursts of a dwarf nova. Most of the dwarf novae with orbital periods shorter than the 2 to 3 hour "period gap" in the orbital distribution show superoutbursts, which last longer and are brighter than normal outbursts. These are referred to as SU UMa dwarf novae after the prototype star. During superoutbursts, the light curves of these CVs display photometric modulations on a period slightly longer than the orbital period. These are known as `superhumps' and result from the dynamical interaction between the disc and the donor star. Although we observed hump-like features in the light curve of the 2009 and 2014 outbursts, they were not the regular saw-tooth modulations typical of superhumps. Moreover both outbursts were short in duration and neither had the prolonged plateau following maximum that would be typical of a superoutburst of an SU UMa dwarf nova. Although we have not observed a superoutburst we cannot yet rule out 1RXS J140429.5+172352 as a member of the SU UMa family since we might have missed one or more superoutbursts due to the observational gaps in our nine year coverage. Alternatively it could be that the superoutburst frequency is low. As yet no orbital period of the system has been published. It might be possible to determine the orbital period via time resolved spectroscopy in quiescence which would help in the classification.

Another possibility is that 1RXS J140429.5+172352 is an Intermediate Polar. This classification is supported by the fact that it is a strong X-ray source and its spectrum is suggestive of a weakly magnetic system (21). The large amplitude and rapid hump-like variations during the outbursts are reminiscent of the flaring or Quasi-Periodic Oscillations (QPOs) seen in several IPs, such as EX Hyi during its 1987 outburst (22). Whilst none of these properties are unique to IPs, we favour this classification based on the currently available evidence.

We note that the outburst behaviour of 1RXS J140429.5+172352 is similar to that of the CV HW Boo in that both show short outbursts in which the outburst light curve is dominated by hump-like variations on the 15 min to 1 hour timescale. HW Boo has $P_{orb}$ = 92.66(17) min (23) which places it well below the period gap, although no superoutbursts have been observed so far. HW Boo might yet be an SU UMa system, although an IP classification is preferred (24). HW Boo also shows a long-term evolution in quiescence brightness over several years similar to what we suspect for 1RXS J140429.5+172352.

Only further observations will resolve the classification of 1RXS J140429.5+172352. Long-term monitoring will be important to determine its true outburst frequency and to identify a potential superoutburst. It is noteworthy that none of the four outbursts reported in this paper was detected by CRTS which highlights the value of amateur astronomers' observations, even in the era of wide-field transient surveys, not only for follow-up observations, but also for discovery of transient events. Detection of a superoutburst should trigger time resolved photometry to look for superhumps, which would be diagnostic of its SU UMa identity. Amateur astronomers, suitably equipped, are ideally placed to contribute to these activities. Similarly further work, including polarimetry to look for circular polarisation and X-ray observations to look for X-ray pulses, may shed light on the possible IP explanation.

Finally we note some ambiguity about the identity of the star in quiescence. Denisenko (12) pointed out that SDSS images of the star in quiescence show a faint blue companion about 0.75 arcsec to the north-east (Figure 5). However, the SDSS spectrum obviously refers to the brightest object so that one should be the CV. Our photometry and astrometry (as well as all the other photometry used in this paper) takes the combined brightness of both stars. If this blue star, which is perhaps magnitude 18 to 19, is the real dwarf nova it would increase the outburst amplitudes reported here by 1 to 2 magnitudes. We attempted to resolve the two stars in our images, both in outburst and quiescence, without success. Moreover comparison of outburst and quiescent astrometry did not yield a significant difference in position. We suggest the matter could be resolved (literally!) by images taken with a much greater plate scale than we currently have access to.

**Conclusions**

The nine-year light curve of the CV 1RXS J140429.5+172352 between 2005 April and 2014 June reveals four dwarf nova-like outbursts. The outbursts amplitude was 3.1 to 5.3 magnitudes above quiescence with a duration of 3 to 5 days. Time resolved photometry during two outbursts revealed small hump-like structures in the light curve with a peak-to-peak amplitude of ~0.5 mag; they were variable in size and occurred on timescales of 20 mins to 1 hour, although without a stable period. Whilst 1RXS J140429.5+172352 has some properties consistent with dwarf novae of the SU UMa family, we suggest that more likely it is an Intermediate Polar. We encourage further observations to enable it to be classified more precisely.

**Acknowledgments**

The authors gratefully acknowledge the use of observations from the AAVSO International Database contributed by observers worldwide. This research also made use of data from the Catalina Real-Time Transient Survey, the MASTER-Amur survey, the All-Sky Automated Survey for Supernovae, the All Sky Automated Survey (ASAS-3), the Sloan Digital Sky Survey, the US Naval Observatory CCD Astrographic Catalog (UCAC2), the ROSAT All-Sky Survey Faint Source Catalogue, and the AAVSO Photometric All-Sky Survey. We also used SIMBAD and Vizier, operated through the Centre de Données Astronomiques (Strasbourg, France) and the NASA/Smithsonian Astrophysics Data System.

JS thanks the Department of Cybernetics at the University of Bradford, UK, for the use of the Bradford Robotic Telescope (BRT), located at the Teide Observatory on Tenerife in the Canary Islands, as part of his monitoring of 1RXS J140429.5+172352 for outbursts.

**Addresses**

Shears: "Pemberton", School Lane, Bunbury, Tarporley, Cheshire, CW6 9NR, UK [bunburyobservatory@hotmail.com]

Dubovský: Vihorlat Observatory Humenne, Slovakia [var@kozmos.sk]

Littlefield: Law School, University of Notre Dame, Notre Dame, IN 46556, USA [clittlef@alumni.nd.edu]

Miller: Furzehill House, Ilston, Swansea, SA2 7LE, UK [furzehillobservatory@hotmail.com]

Morelle: Lauwin-Planque Observatory, F-59553 Lauwin-Planque, France [etmor@free.fr]

Pickard: 3 The Birches, Shobdon, Leominster, Herefordshire. HR6 9NG, UK [roger.pickard@sky.com]

Ruiz: Observatorio de Cantabria, Agrupacion Astronomica Cantabra, Instituto de Fisica de Cantabria (CSIC-UC), Avda Los Castros s/n, 39005 Santander, Cantabria, Spain [parhelio@astrocantabria.org]

Sabo: 2336 Trailcrest Drive, Bozeman, MT 59718, USA [rsabo333@gmail.com]

| Observer | Telescope | CCD |
|---|---|---|
| Dubovský | 0.25 m reflector | Meade DSI Pro |
| Littlefield | 0.28 m SCT | SBIG ST-8XME |
| Miller | 0.35 m SCT | Starlight Xpress SXVR-H16 |
| Morelle | 0.4 m SCT | SBIG ST9 |
| Pickard | 0.35 m SCT | Starlight Xpress SXVF-H9 |
| Ruiz | 0.4 m SCT | SBIG ST-8XME |
| Sabo | 0.43 m reflector | SBIG STL-1001 |

**Table 1: Instrumentation used for time resolved photometry**

| Date (UT) | Start time (JD) | End time (JD) | Duration (h) | Observer |
|---|---|---|---|---|
| **2009** | | | | |
| May 2 | 2454953.692 | 2454953.875 | 4.0 | Sabo |
| May 2 | 2454954.357 | 2454954.615 | 6.2 | Morelle |
| May 2 | 2454954.429 | 2454954.581 | 5.4 | Dubovský |
| May 3 | 2454955.317 | 2454955.577 | 6.2 | Dubovský |
| May 4 | 2454956.312 | 2454956.341 | 0.7 | Dubovský |
| **2014** | | | | |
| May 17 | 2456795.486 | 2456795.575 | 2.1 | Ruiz |
| May 19 | 2456796.757 | 2456796.876 | 2.9 | Littlefield |
| May 20 | 2456797.814 | 2456797.925 | 2.7 | Sabo |
| May 20 | 2456798.389 | 2456798.574 | 4.4 | Pickard |
| May 20 | 2456798.464 | 2456798.582 | 2.8 | Miller |
| May 21 | 2456798.708 | 2456798.921 | 5.1 | Sabo |
| May 22 | 2456799.604 | 2456799.841 | 5.7 | Littlefield |
| May 25 | 2456802.692 | 2456802.780 | 2.1 | Sabo |

**Table 2: Log of time resolved photometry**

| Date of outburst | JD | Maximum brightness (mag) | Duration (d) | Time since previous outburst (d) |
|---|---|---|---|---|
| 2008 Jun 27 | 2454644 | 11.9 | ND | |
| 2009 May 1 | 2454953 | 13.1 | ~3 | 309 |
| 2012 Jan 26 | 2455953 | 14.2 | ND | 1000 |
| 2014 May 17 | 2456794 | 14.1 | ~5 | 841 |

**Table 3: Outbursts of 1RXS J140429.5+172352 recorded between 2005 April and 2014 June**

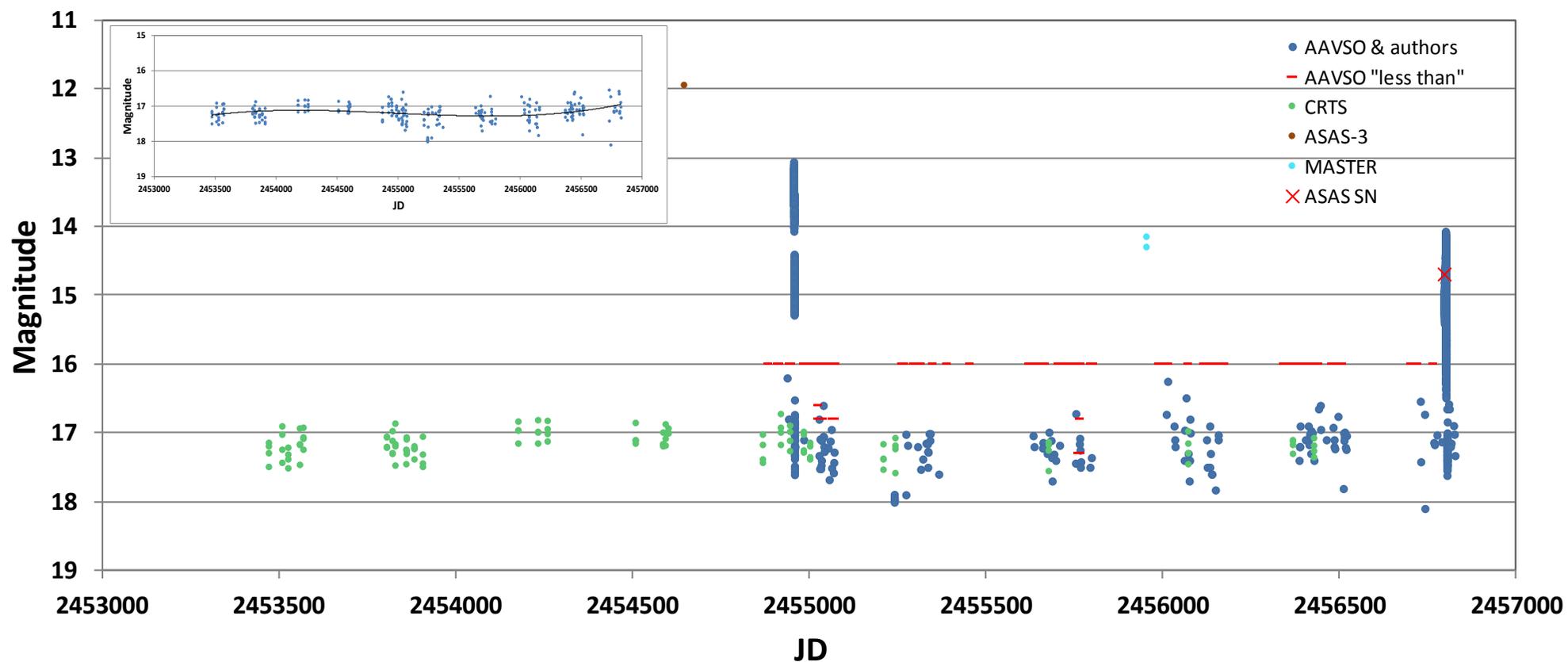

Figure 1: The nine year light curve of 1RXS J140429.5+172352, from 2005 April to 2014 July. Inset is a plot showing the same data, having removed all outbursts and "fainter than" data, to which a third order polynomial has be fitted

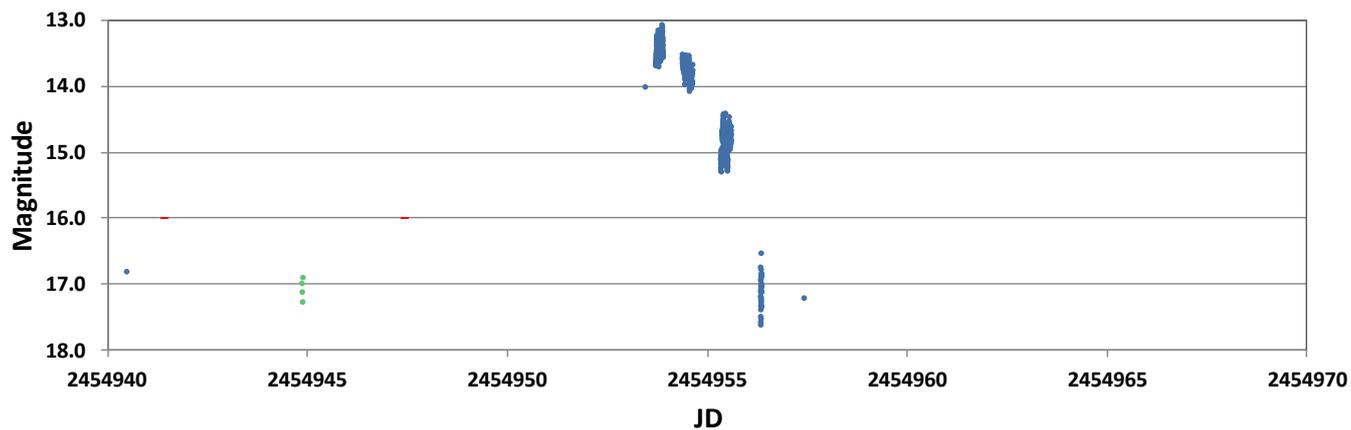

**2009**

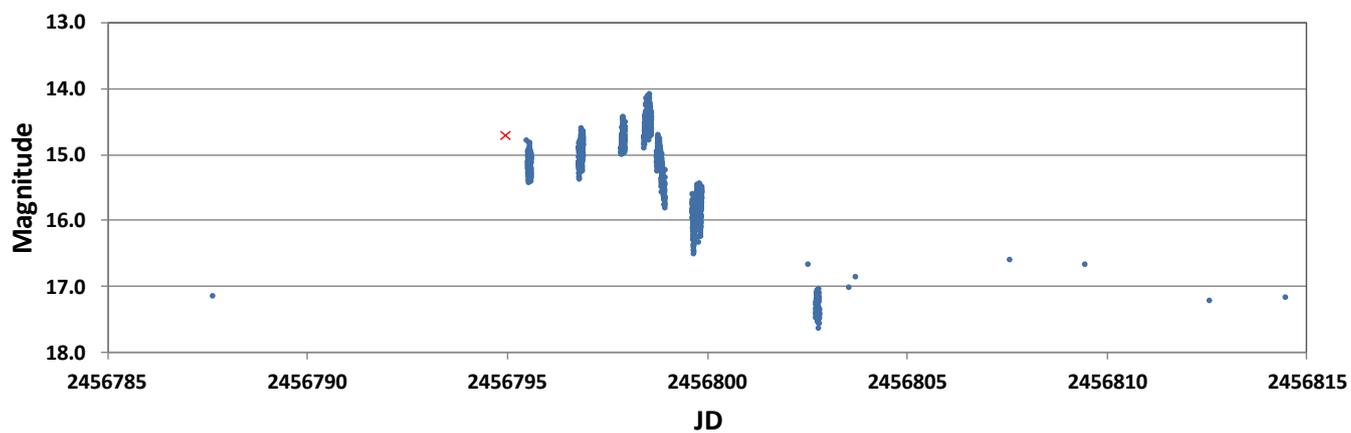

**2014**

**Figure 2: Light curves of the 2009 (top) and 2014 (bottom) outbursts**

See Figure 1 for key to symbols

*Accepted for publication in the Journal of the British Astronomical Association*

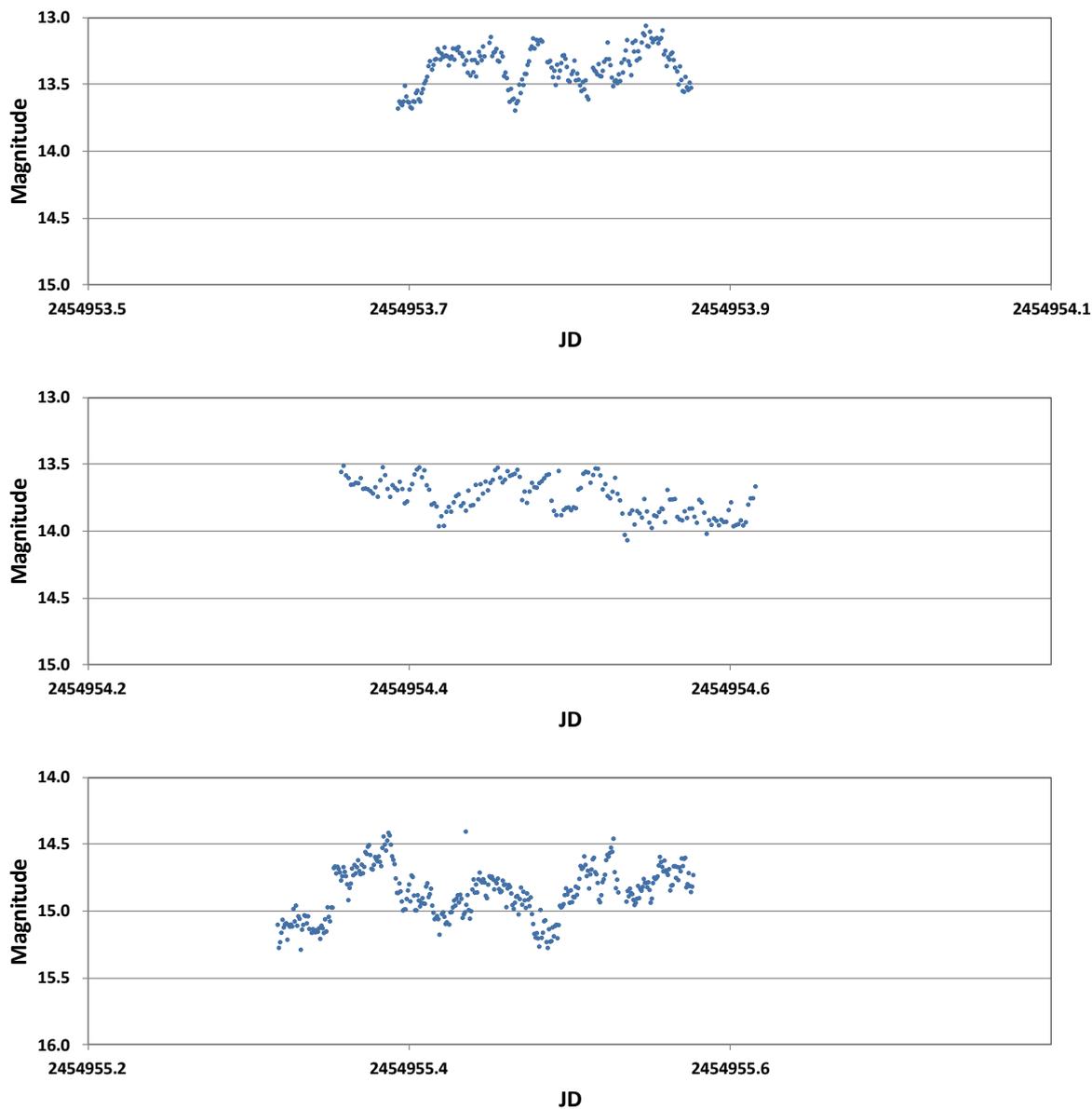

**Figure 3: Time resolved photometry during the 2009 outburst**

---

**Figure 4: Time resolved photometry during the 2014 outburst**

**(on following page)**

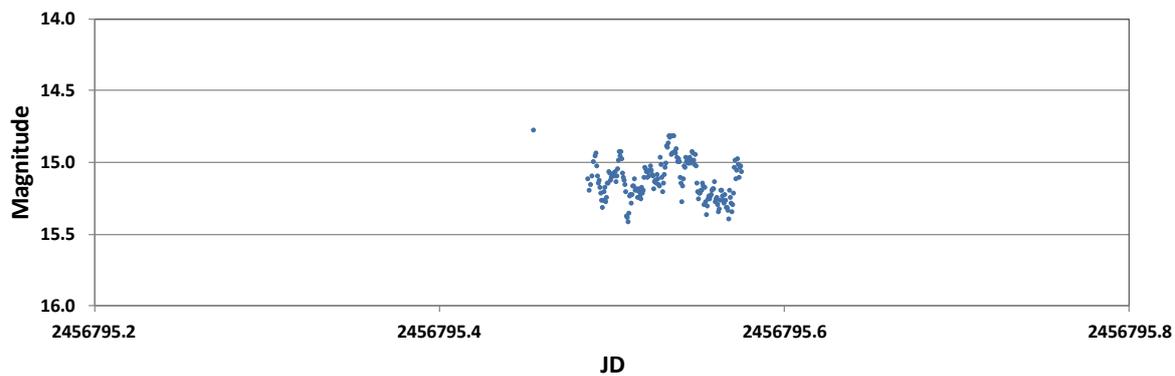

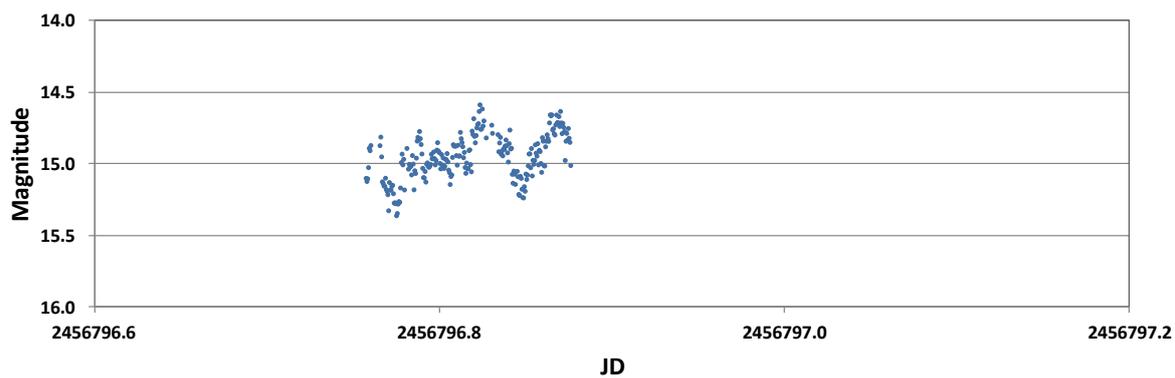

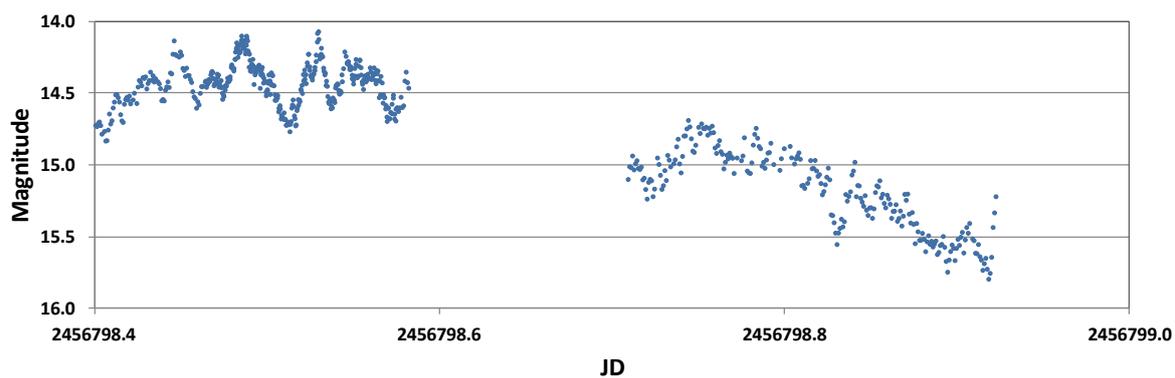

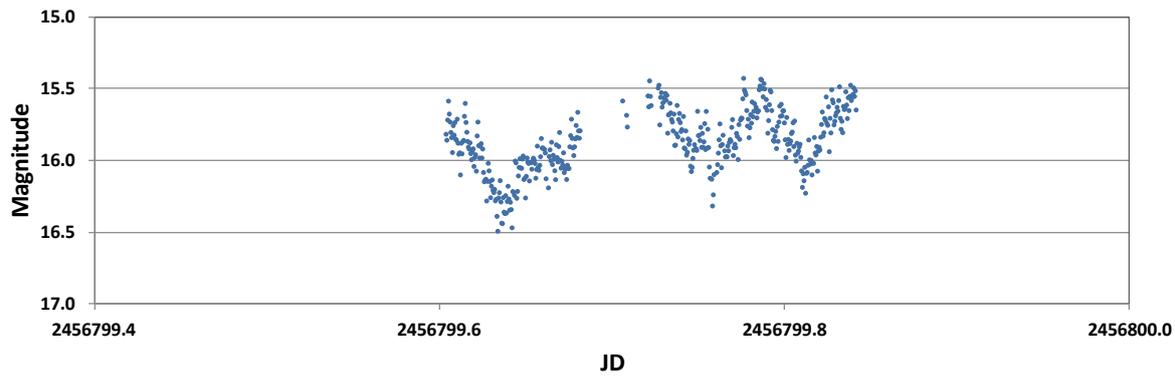

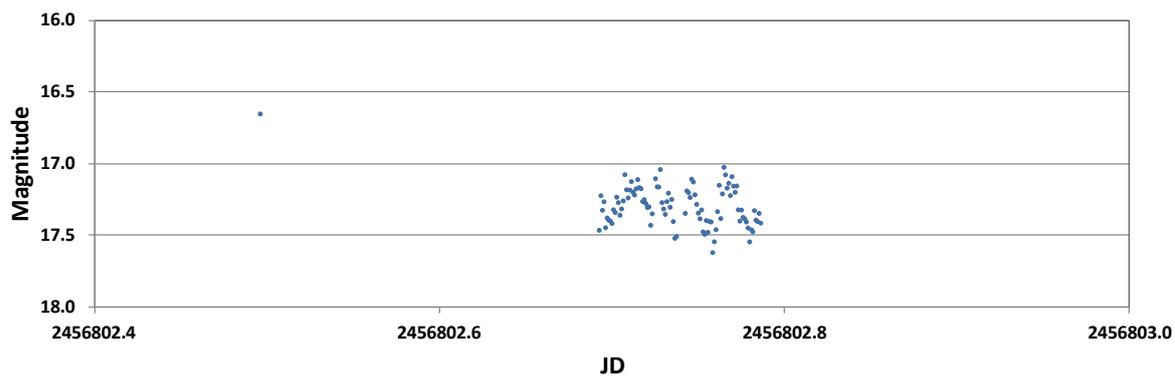

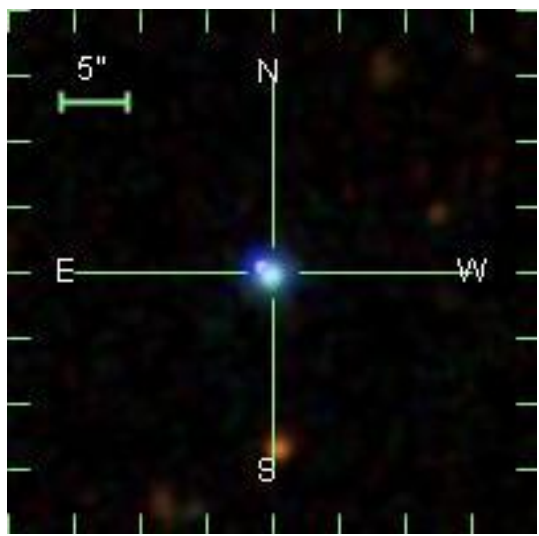

**Figure 5: SDSS image of SDSS J140429.37+172359.4 showing a faint blue companion to the NE** (5)